# The Phase Transition to the Quark Gluon Plasma: Recent Results from Lattice Calculations

F. Karsch[a] *

[a]Fakultät für Physik, Universität Bielefeld,
Postfach 100131, D-33501 Bielefeld, Germany.

We will discuss here some of the recent results obtained from lattice simulations of QCD at non-zero temperature. Such calculations aim at a quantitative understanding of the thermodynamics of strongly interacting matter in equilibrium. We concentrate on a discussion of the equation of state, the chiral transition in two-flavour QCD and hadronic properties related to chiral symmetry restoration in the vicinity of $T_c$.

## 1. INTRODUCTION

One of the most exciting predictions of QCD is the existence of a phase transition at some critical temperature to a new phase of strongly interacting matter – the quark gluon plasma (QGP). Lattice calculations aim at a quantitative understanding of this phase transition and a determination of the equilibrium properties of both phases of strongly interacting matter.

A well established result from lattice calculations is the strong dependence of the order of the phase transition on the QCD symmetries related to the colour as well as the flavour degrees of freedom. In the absence of quarks, ie. in pure $SU(N)$ gauge theories, the phase transition is second order for $N = 2$ and first order for $N = 3$. In the case of QCD with $n_f$ light quark flavours the transition is first order for $n_f \geq 3$ and does seem to be continuous for $n_f = 2$. The lattice calculations also show a strong dependence of the transition temperature on the number of partonic degrees of freedom. The transition temperature, expressed for instance in units of the square root of the string tension ($T_c/\sqrt{\sigma}$), is substantially smaller in QCD with light quarks than in a pure gauge theory. In the case of QCD with two light quarks one finds $T_c(n_f = 2) \simeq 150$MeV, while in the purely gluonic theory, $n_f \equiv 0$, $T_c$ is found to be substantially higher, $T_c(n_f = 0) \simeq 260$MeV. This may be understood in terms of a percolation threshold: In the absence of quarks there are no light hadrons, the lightest states are glueballs with a mass of $O(1$ GeV$)$. Therefore, a rather high temperature is needed to excite enough of these heavy glueball states in order to reach a critical density where hadrons start overlapping.

A central goal of lattice calculations at finite temperature is to reach a quantitative understanding of the equation of state for strongly interacting matter and to learn details

*Talk given at XI. International Conference on Ultra-Relativistic Nucleus-Nucleus-Collisions, January 9-11, 1995, Monterey, USA. We gratefully acknowledge support through the Deutsche Forschungsgemeinschaft under contract No. Pe 340/3-2

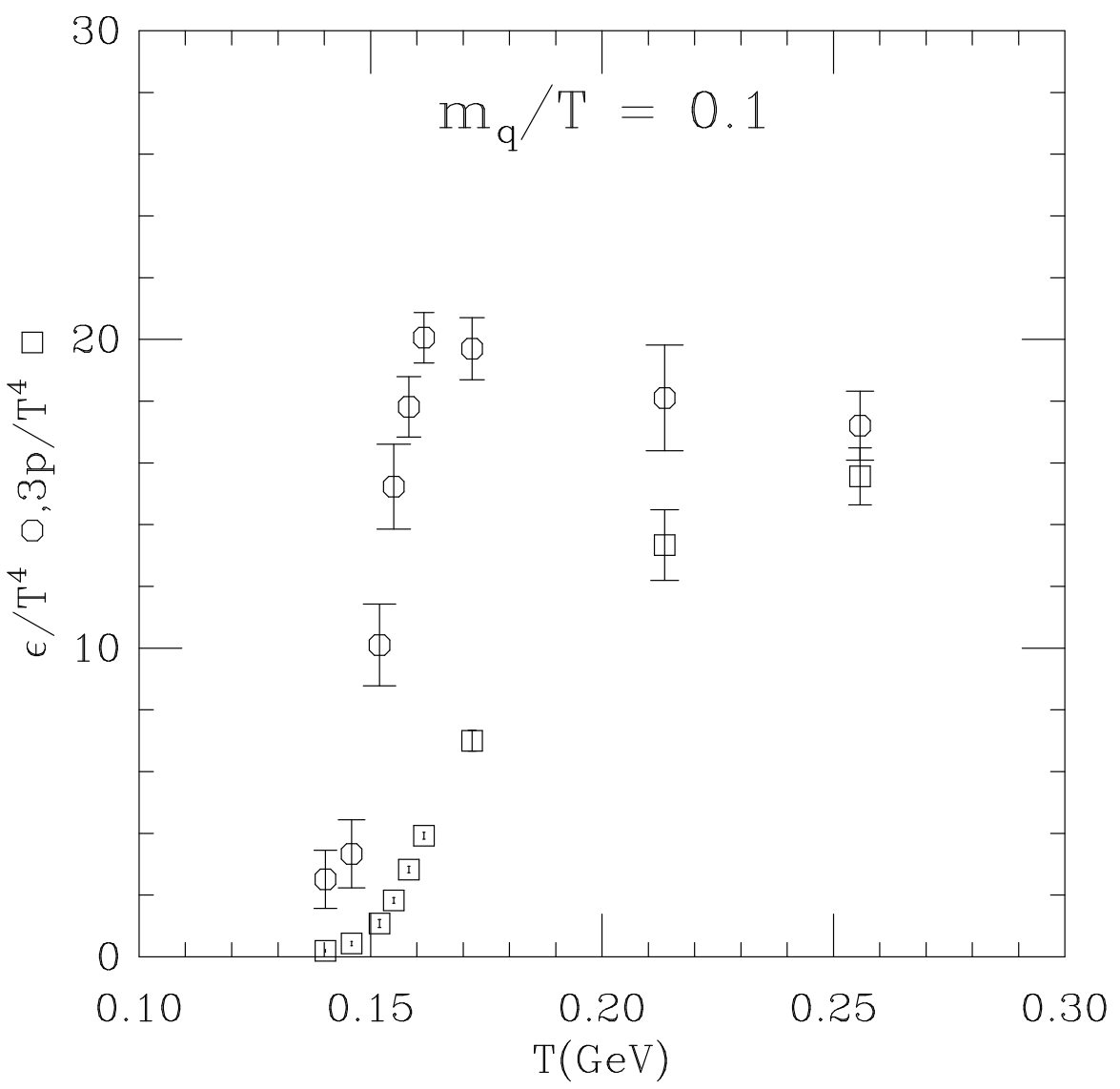

Figure 1. Energy density and pressure for two-flavour QCD on lattices with temporal extent $N_\tau = 4$.

about its structure in the vicinity of $T_c$, a regime which is not accessible to perturbative methods. The theoretical approach to a truly non-perturbative calculation of bulk thermodynamic quantities has improved a lot over recent years. In Figure 1 we show results from a recent non-perturbative calculation of the energy density and pressure in two-flavour QCD [1]. The rapid change of the energy density in a narrow region around $T_c \sim 150$MeV and the comparatively slow rise of the pressure is clearly seen. A more detailed discussion of the equation of state, in particular the approach to the high temperature limit and its relation to perturbative calculations is, however, hardly possible with the present results for QCD with light dynamical quarks. The finite size effects are still quite large on the lattices presently accessible to simulations with dynamical quarks. Such effects have to be controlled before a comparisons with perturbative calculations in high temperature QCD will be possible. In the next section we will therefore restrict ourselves to a discussion of the equation of state in gauge theories, where a systematic analysis of the size dependence of lattice results is now possible [2,3]. In section 3 we will discuss results of studies of the order of the phase transition in two flavour QCD. A discussion of the influence of the chiral transition on hadron masses and other hadronic properties is presented in section 4. Finally we give our conclusions in section 5.

## 2. EQUATION OF STATE

The calculation of energy density and pressure in finite temperature QCD has been a central target of lattice calculations. In particular, the direct calculation of the free energy density [4] combined with our knowledge about contributions of non-perturbative effects to the lattice QCD $\beta$-function [5] allowed to eliminate remaining perturbative elements which were still present in earlier approaches. Nonetheless, up to very recently lattice calculations of bulk thermodynamic quantities have been restricted to lattices with a rather short extent in the temporal direction [2]. In fact, all existing results for the equation of state of QCD with dynamical fermions, still are restricted to lattices of size $N_\sigma^3 \times N_\tau$ with $N_\tau = 4$.

In the following we are going to discuss a systematic analysis of the lattice equation of state (EOS) for pure gauge theories and its extrapolation to the continuum limit. We will discuss the importance of finite lattice effects for the high temperature limit of the EOS as well as in the vicinity of $T_c$. This may be viewed as a test case for future systematic studies of the EOS of QCD with dynamical fermions.

### 2.1. The high temperature limit

The standard forms of the lattice discretized gluon and fermion actions are known to lead to a strong cut-off (lattice spacing $a$) dependence of thermodynamic quantities on finite lattices of size $N_\sigma^3 \times N_\tau$. At finite temperature these ultraviolet cut-off effects are controlled by $aT \equiv 1/N_\tau$. Therefore they do not disappear when the "continuum" limit is taken at fixed $N_\tau$, ie. when the $(a \rightarrow 0)$-limit is mixed with the $(T \rightarrow \infty)$-limit. These cut-off effects decrease with increasing $N_\tau$ like $(aT)^2 \equiv 1/N_\tau^2$. For the gluonic part of the energy density of an ideal gluon gas ($\epsilon_{\rm SB}$) one has, for instance, [2]

$$\epsilon_a(N_\tau) \equiv \left(\frac{\epsilon_{\rm SB}}{T^4}\right)(N_\tau) = (N^2-1)\left[\frac{\pi^2}{15} + \frac{2\pi^4}{63}\cdot (aT)^2 + O\Big((aT)^4\Big)\right] \quad , \tag{1}$$

where $N$ denotes the number of colours. On small lattices also the higher order corrections are still important. For instance, on a lattice with temporal extent $N_\tau = 4$ finite size effects lead to a 50% distortion of the free gluon gas, ie. $\epsilon_a(4)$ is 50% larger than the corresponding continuum Stephan-Boltzmann constant. On a twice as large lattice this reduces to less than 10%, $\epsilon_a(8) \simeq 1.07$. It is well known from calculations on lattices with $N_\tau = 4$ that, for instance, the energy density above $T_c$ rapidly approaches this "distorted" ideal gas value. However, in how far perturbative and/or non-perturbative contributions to the energy density are hidden under the large finite size corrections can only be clarified through a systematic analysis on various size lattices.

In the pure gauge sector, ie. for $SU(2)$ [2] and $SU(3)$ [3] gauge theories, we have recently performed detailed calculations of thermodynamic quantities on lattices with temporal extent $N_\tau = 4$, 6 and 8. In Figures 2a and 2b we show results for the energy density and pressure in a $SU(3)$ gauge theory. These have been obtained from calculations on lattices of size $16^3 \times 4$ and $32^3 \times N_\tau$ with $N_\tau = 6$ and 8. As can be seen the finite cut-off effects visible in the calculation of the pressure and energy density are similar to those calculated in the ideal gas limit. We note that the finite size effects are largest for high temperature. This is to be expected as the finite cut-off mainly affects the high momentum modes, which dominate the energy density and pressure in that limit. Still on a quantitative level these effects are smaller than expected on the basis of the perturbative calculation for an ideal gas. This suggest that low momentum modes, which are not sensitive to the short distance cut-off effects, still provide a relatively large contribution to the energy density and pressure.

We can use the leading $1/N_\tau^2$ dependence of energy density and pressure for an extrapolation of the $N_\tau = 6$ and 8 results shown in Figure 2 to the continuum limit. This suggests that already the $N_\tau = 8$ results themselves provide a very good approximation to the continuum limit.

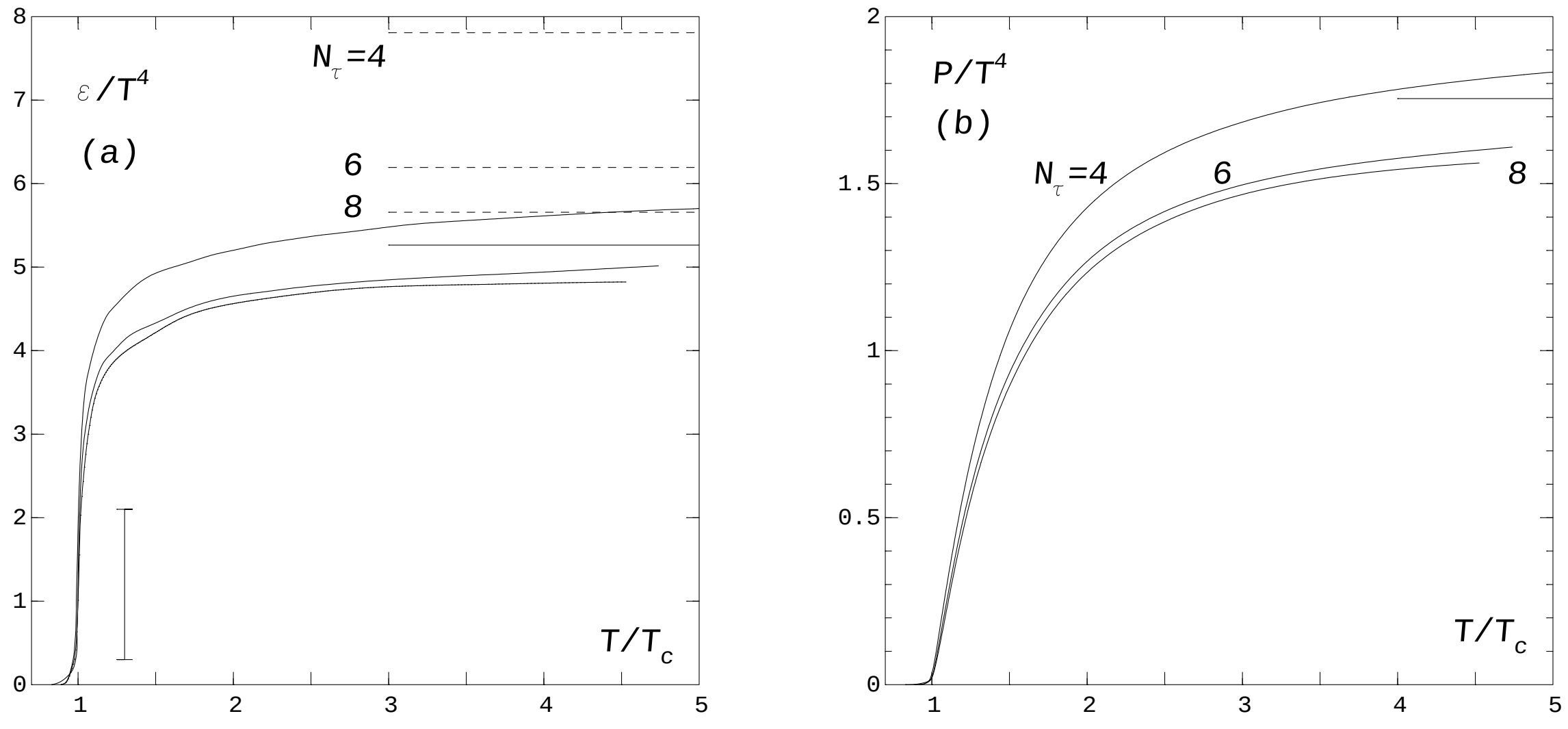

Figure 2. Energy density (a) and pressure (b) in a $SU(3)$ gauge theory on $16^3\times 4$, $32^3\times 6$ and $32^3\times 8$ lattices. The dashed horizontal lines indicate the corresponding result for an ideal gas on these lattices. The solid horizontal line shows the continuum ideal gas result. The vertical bar in (a) gives the size of the latent heat discontinuity at $T_c$ [6].

We note that the increase in pressure is much slower compared to the that in the energy density. At $T = 2T_c$ the energy density deviates from the ideal gas value only by about 15% while the pressure is still more than 30% below the Stephan-Boltzmann value. This has strong consequences for the velocity of sound, which deviates in this temperature interval above $T_c$ strongly from the ideal gas value, $c_s^2 = 1/3$ (Figure 3).

The complicated infrared structure of QCD is expected to play an important role also for the thermodynamics at high temperature. Non-perturbative modifications of the low momentum spectrum are expected to be responsible for a substantial part of the deviations of the EOS from ideal gas behaviour, $\epsilon = 3p$. The leading ideal gas term is eliminated in the difference $\epsilon - 3p$. If infrared physics is mainly responsible for the deviations of this quantity from zero, we should expect that it is less sensitive to finite cut-off effects then $\epsilon$ and $p$ separately. This is indeed the case as can be seen from Figure 4.

The difference $\epsilon - 3p$ is given by the trace of the energy-momentum tensor normalized to its zero temperature value. It thus may be viewed as the definition of the temperature dependent gluon condensate [7]

$$\epsilon - 3p = \langle G^2\rangle_0 - \langle G^2\rangle_T \quad . \tag{2}$$

where $\langle G^2\rangle_0$ denotes the gluon condensate at $T = 0$. Perturbative corrections start at $O(g^4)$,

$$\frac{\epsilon - 3p}{T^4} = -T\frac{\mathrm{d}g(T)}{\mathrm{d}T}\left[\frac{1}{144}N(N^2-1)g(T) + O(g^2)\right] \quad . \tag{3}$$

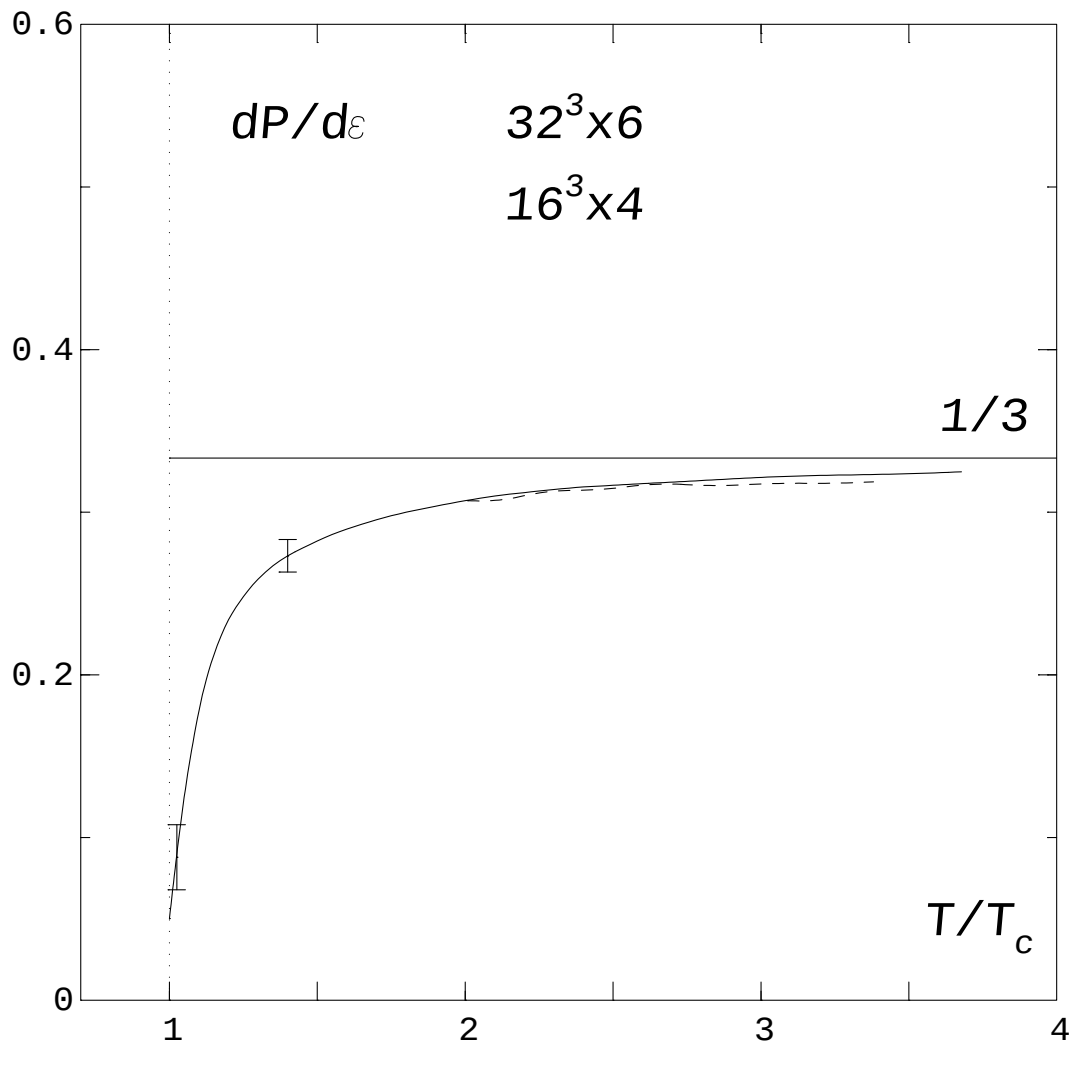

Figure 3. The velocity of sound in a $SU(3)$ gauge theory. Results for the $32^3 \times 8$ lattice are indistinguishable from those shown for the $32^3 \times 6$ lattice.

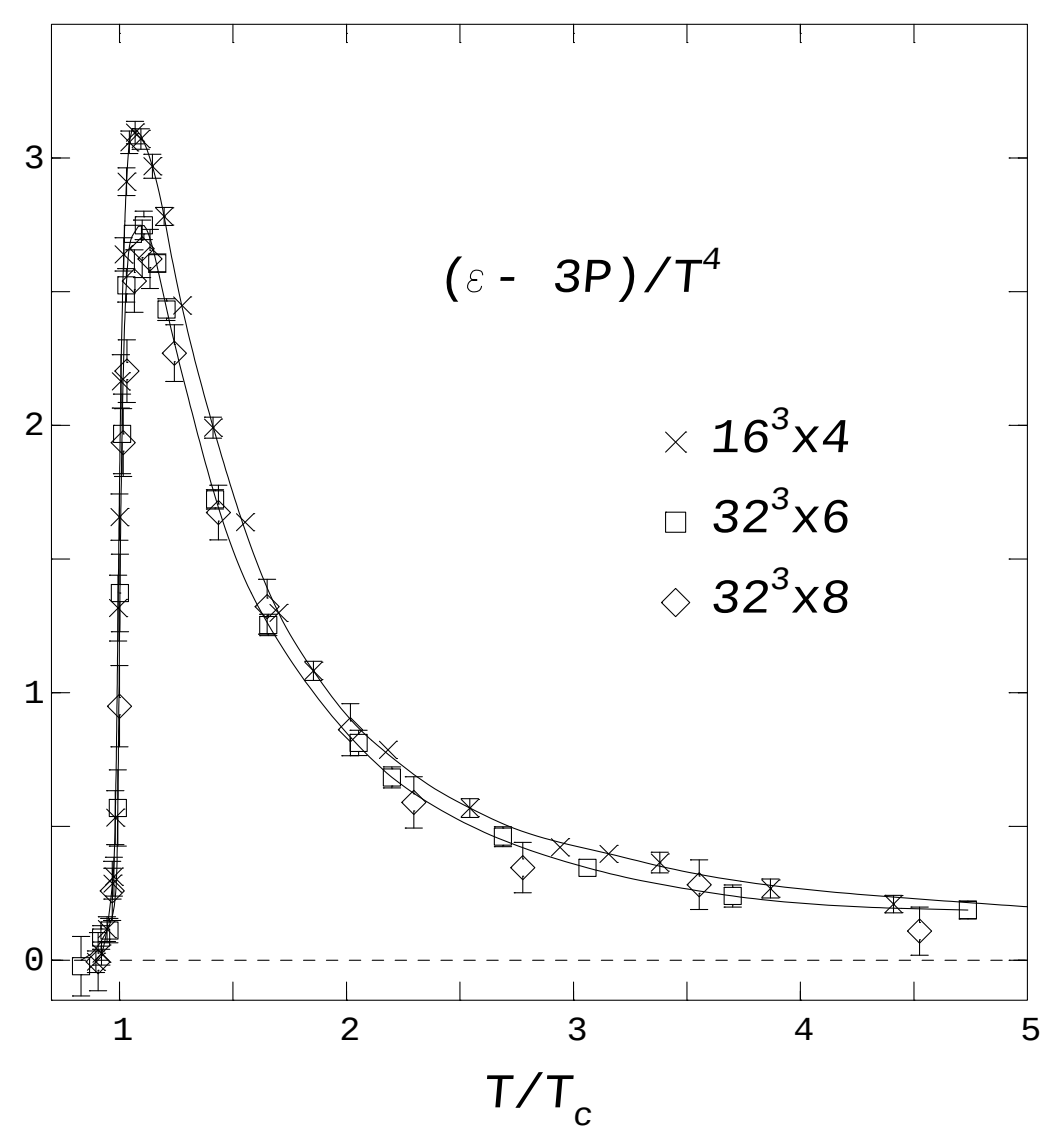

Figure 4. The interaction measure $\epsilon - 3p$ normalized to $T^4$ for a $SU(3)$ gauge theory on three different lattices.

The difference $(\epsilon - 3p)/T^4$ thus is expected to approach zero only logarithmically with increasing temperature. From the high temperature behaviour shown in Figure 4 we find that $g(T) \gtrsim 1$ is needed to describe this regime with the leading order perturbative correction. Unfortunately, the perturbative expansion cannot be trusted for such large values of $g(T)$ [8].

In order to express $(\epsilon - 3p)$ in physical units and relate the numerical results to the zero temperature estimates for the gluon condensate we have to set the scale for the temperature. This can, for instance, be achieved by using results from lattice QCD for the ratio $T_c/\sqrt{\sigma}$ with $\sqrt{\sigma} \simeq 420$MeV. In order to do so we have re-calculated the critical couplings for the deconfinement transition on $32^3 \times N_\tau$ lattices with $N_\tau = 6$, 8 and 12. At these couplings we also have calculated the string tension on lattices of size $32^4$. From this we find $T_c/\sqrt{\sigma} \simeq 0.62$ [3][2].

$(\epsilon - 3p)/T^4$ shows a pronounced peak close to $T/T_c = 1.1$. Using the result for $T_c/\sqrt{\sigma}$ we deduce from Figure 4

$$\frac{\epsilon - 3p}{T^4} \simeq 2.4 \frac{\mathrm{GeV}}{\mathrm{fm}^3} \quad \mathrm{at} \quad T = 1.1 T_c \;, \tag{4}$$

which should be compared with estimates for the gluon condensate at zero temperature, $\langle G^2 \rangle_0 = 2.0$ Gev/fm$^3$ [7]. This suggests that the constant zero-temperature gluon condensate gives the dominant contribution to $(\epsilon - 3p)/T^4$ close to $T_c$, ie. $\langle G^2 \rangle_T$ turns out to be small at $T = 1.1T_c$. It becomes negative in this temperature region and approaches the negative perturbative behaviour indicated by Eq. 3.

[2]This is about 10% larger than our earlier estimate [9], which was based on old results for the critical couplings calculated on rather small spatial lattices.

### 2.2. The critical energy density

In the vicinity of the critical temperature the dependence of the energy density on $N_\tau$ itself is small. Standard renormalization group arguments (finite size scaling) suggest, that in this region the dominant size dependence is not related to the non-vanishing lattice spacing but rather to the large correlation length close to $T_c$, ie. the low momentum structure of the free energy. In fact, the behaviour of thermodynamic quantities in this region is controlled by the singular part of the free energy density,

$$f_s(t,V) = \frac{1}{VT^3} Q\left(t(TV^{1/3})^{1/\nu}\right) \quad , \tag{5}$$

where $t = (T - T_c)/T_c$ denotes the reduced temperature, $Q$ is an arbitrary scaling function and $\nu$ denotes the correlation length critical exponent. In terms of the lattice parameters $N_\sigma$ and $N_\tau$ we have $TV^{1/3} = N_\sigma/N_\tau$. Finite size effects in the vicinity of $T_c$ are thus controlled by this ratio rather than by $aT = 1/N_\tau$ alone. The scaling behaviour in terms of $TV^{1/3}$ has clearly been seen in the case of an $SU(2)$ gauge theory and could be used there to extract the critical energy density [2] in the thermodynamic limit. The $SU(2)$ gauge theory has a second order deconfinement phase transition. The critical behaviour is controlled by critical exponents of the 3-$d$ Ising model. One thus expects the critical energy density to scale like

$$\frac{\epsilon_c}{T_c^4} = \left(\frac{\epsilon_c}{T_c^4}\right)_\infty + \text{const.}\ (TV^{1/3})^{(\alpha-1)/\nu} \quad , \tag{6}$$

where $\alpha$ and $\nu$ denote the critical indices of the $3d$ Ising model, $(\alpha - 1)/\nu = -1.41$. As can be seen from Figure 5 this scaling behaviour holds very well. We note that the critical energy density turns out to be rather small, $(\epsilon_c/T_c^4) = 0.256(23)$ or $\epsilon_c/\epsilon_{\rm SB} \simeq 1/8$.

A similarly detailed analysis of the finite volume effects close to $T_c$ does not exist for the energy density of the $SU(3)$ gauge theory. However, the latent heat at the phase transition point has been studied on fairly large lattices [6]. The experience from the $SU(2)$ analysis suggest that the energy density at $T_c$ in the plasma phase, $\epsilon^+(T_c)$, provides a good estimate for the critical energy density. From Figure 2 we find $\epsilon^+(T_c)/T_c^4 \simeq 2$. Using again the string tension to set a physical scale we find here a critical energy density of about 1 GeV/fm$^3$. A similar value is also obtained from the energy density calculation for two-flavour QCD shown in Figure 1. We deduce from this Figure $\epsilon_c/T_c^4 = 6 \pm 4$. With a critical temperature of 150 MeV this also yields a rather small critical energy density,

$$\epsilon_c \leq 1\ \text{GeV/fm}^3 \ . \tag{7}$$

We note, however, that this estimate (as usual) is rather sensitive to the value of $T_c$ and might still change with more accurate determinations of the chiral transition temperature.

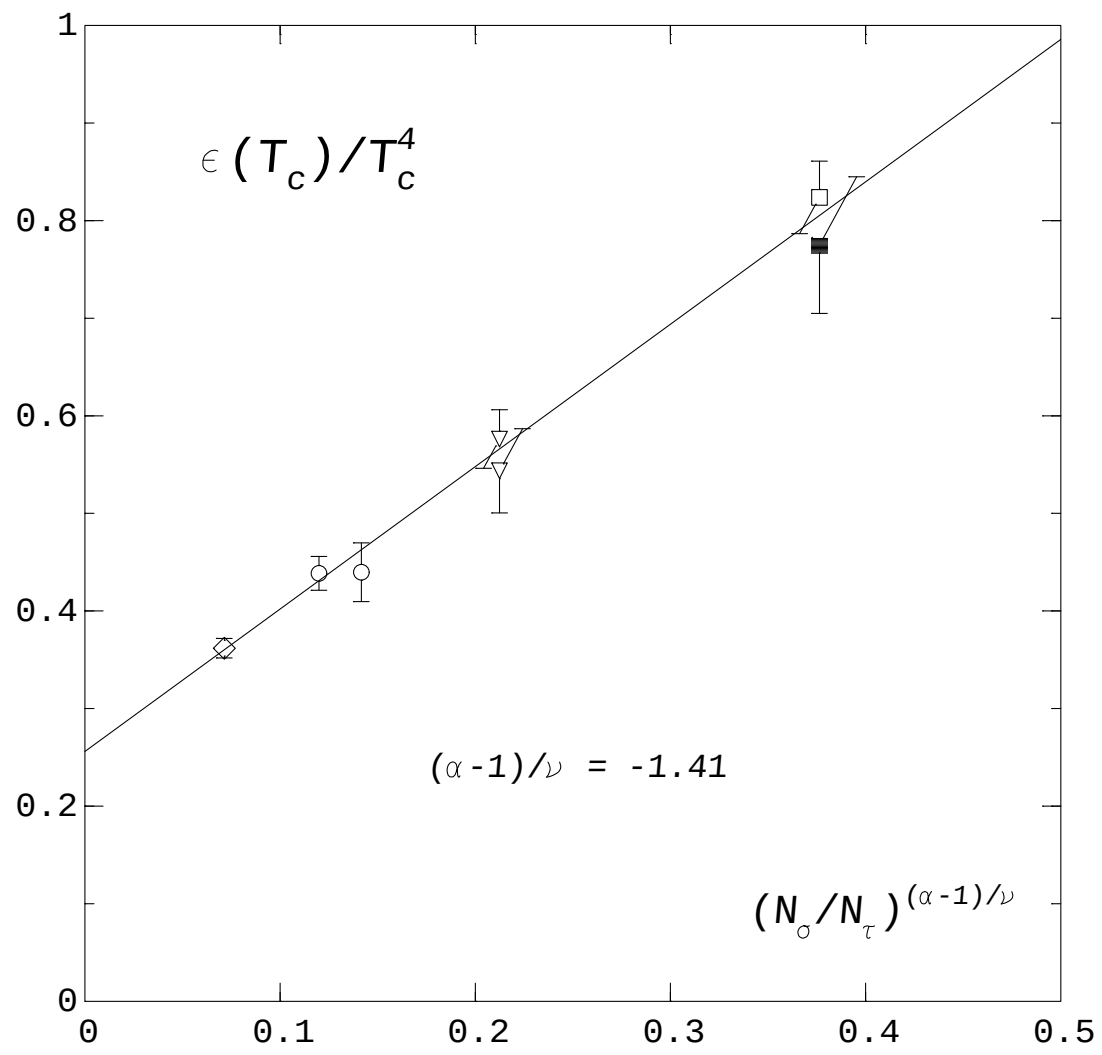

Figure 5. Extrapolation of the critical energy density for the $SU(2)$ gauge theory as function of $TV^{1/3} = N_\sigma/N_\tau$ to the infinite volume limit. Shown is a collection of data from various lattice sizes.

## 3. THE CHIRAL PHASE TRANSITION

The nature of the chiral phase transition does seem to depend strongly on the number of light quark flavours, $n_f$. Monte Carlo calculations performed so far with two light flavours suggest that the transition is continuous, while it seems to be first order for $n_f \geq 3$. The special role of two-flavour QCD has been noticed already some time ago by Pisarski and Wilczek [10]. They have suggested that the dynamics of the chiral phase transition is controlled by an effective, 3-$d$ scalar Lagrangian constructed in terms of the chiral order parameter field, $\sigma \sim \bar{\psi}\psi$. This effective theory has a global $O(4)$ symmetry ($SU(2) \times SU(2)$), which is spontaneously broken. It therefore is expected that in the case of a continuous transition the critical behaviour is governed by critical exponents of a 3-$d$, $O(4)$ symmetric spin model [10,11]. A remaining question in this scenario is to what extent the axial $U(1)$ influences the critical behaviour in the vicinity of $T_c$ [11,12]. An even more fundamental problem is whether an effective scalar theory can at all describe the critical behaviour of a theory, in which the scalar fields are only constructed as fermionic bilinears. It recently has been argued by Kocić and Kogut that this may not be the case and that two-flavour QCD may show mean field behaviour in the vicinity of $T_c$ [13]. On the lattice an additional problem arises: None of the fermion Lagrangians used for numerical calculations has the complete chiral symmetry of the continuum Lagrangian. In the staggered fermion formulation, for instance, one has a chiral $U(1)$ rather than the full $SU(2)$ symmetry, which only is restored in the continuum limit. This too may influence the critical behaviour in numerical studies. A detailed quantitative investigation of the critical behaviour thus is needed.

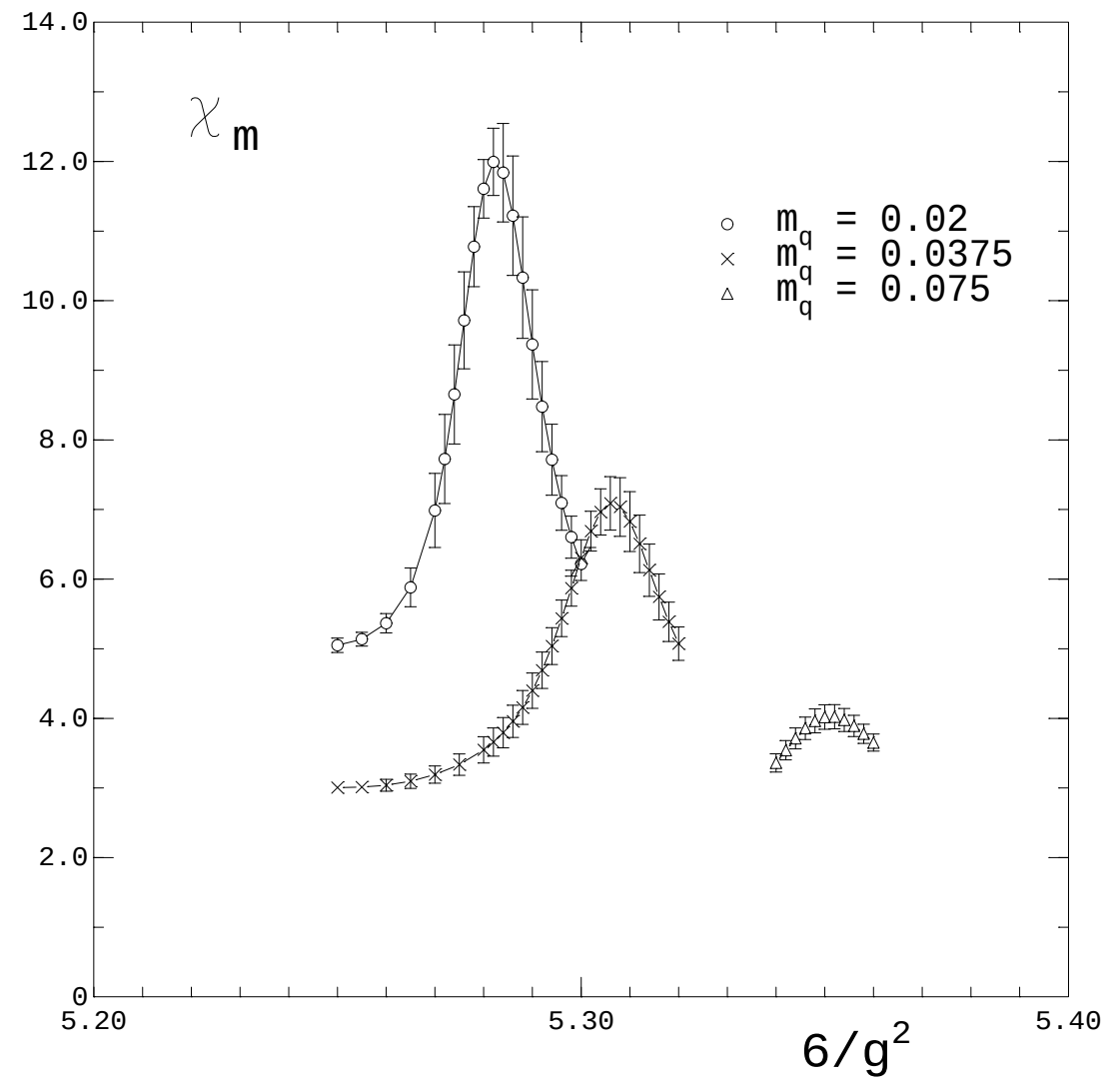

Figure 6. The chiral susceptibility versus $6/g^2$ for three values of the quark mass on a $8^3 \times 4$ lattice.

The critical behaviour of QCD with light quarks is controlled by two dimensionless parameters, the reduced temperature $t = (T - T_c)/T_c$ and the scaled quark mass, $h = m_q/T$. The latter corresponds to an external magnetic field in spin models. In the case of QCD with dynamical quarks it is the quark mass rather than the spatial lattice volume, which is the most stringent limitation for large correlation lengths in the vicinity of the critical point. Rather than performing a finite size scaling analysis, as it has been done in the case of the deconfinement transition in $SU(N)$ gauge theories, it is more appropriate to consider the scaling behaviour in terms of the quark mass. Finite volume effects may in a first approximation be ignored.

In the vicinity of the critical point the behaviour of bulk thermodynamic quantities is related to thermal ($y_t$) and magnetic ($y_h$) critical exponents, which characterize the scaling behaviour of the singular part of the free energy density,

$$f_s(t,h) \equiv -\frac{T}{V} \ln Z_s = b^{-1} f(b^{y_t} t, b^{y_h} h) \quad . \tag{8}$$

Here $b$ is an arbitrary scale factor. Various scaling relations for thermodynamic quantities can be derived from Eq.8 [14,15]. For instance, one finds for the chiral order parameter, $\langle \bar\psi\psi \rangle$, and its derivative with respect to the quark mass (the chiral susceptibility $\chi_m$)

$$\begin{aligned} \langle \bar\psi\psi \rangle(t,h) &= h^{1/\delta} F(z) \\ \chi_m(t,h) &= \frac{1}{\delta} h^{1/\delta - 1} \Big[ F(z) - \frac{z}{\beta} F'(z) \Big] \quad , \end{aligned} \tag{9}$$

with scaling functions $F$ and $F'$ that only depend on a specific combination of the reduced temperature ($t$) and scaled quark mass ($h$), $z = th^{-1/\beta\delta}$. The critical exponents $\beta$ and $\delta$ are given in terms of $y_t$ and $y_h$ as $\beta = (1 - y_h)/y_t$ and $\delta = y_h/(1 - y_h)$. A direct consequence of Eq. 9 is, for instance, that $\chi_m$ has a peak located at some fixed value of $z_c$. This defines a pseudo-critical point, $t_c(h) \equiv z_c h^{1/\beta\delta}$, at which the peak height of $\chi_m$ increases with decreasing quark mass according to Eq. 9. Figure 6 shows the scaling behaviour of $\chi_m$ for three values of the quark mass [15]. The rapid rise of $\chi_m$ with decreasing quark mass can be used to determine the critical exponent $\delta$.

A quite useful observable is the chiral cumulant,

$$\Delta(z) = \frac{h\chi_m}{\langle\bar{\chi}\chi\rangle} \quad . \tag{10}$$

From Eq. 9 it follows that $\Delta$ is directly a scaling function, ie. it only depends on $z$. Moreover, it is uniquely determined at $z = 0$, where it takes on the value $1/\delta$ even at non-vanishing values of the quark mass. The chiral cumulant thus allows a direct calculation of this critical exponent. In Figure 7a we show a collection of the pseudo-critical couplings obtained from Monte Carlo calculations with various values of the quark mass and on various different lattice sizes. The curves show various fits for the quark mass dependence of these pseudo-critical couplings. Apparently it is difficult to distinguish $O(4)$ exponents from mean field exponents in this way. We note, however, that the location of the zero quark mass critical coupling is quite well determined. The cumulant $\Delta$ is shown in Figure 7b. For the two smaller quark masses this quantity has been evaluated in the region of the estimated zero quark mass critical coupling, $\beta_c(m_q = 0) \simeq 5.243(10)$. In this interval the cumulant takes on values between 0.21 and 0.26, which is compatible with the $O(4)$ exponent $1/\delta = 0.205(45)$.

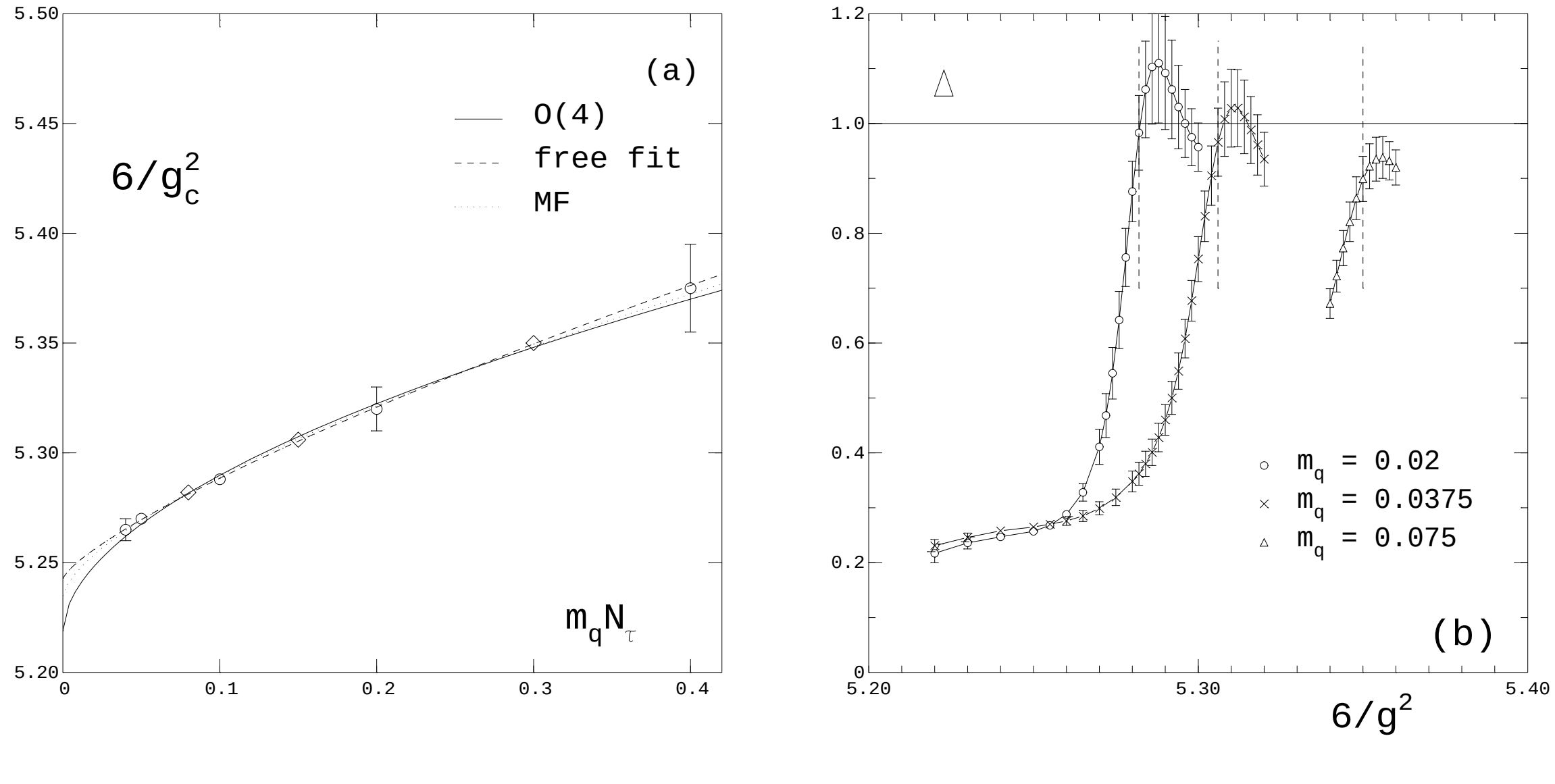

Figure 7. The pseudo-critical couplings (a) of the chiral transition in two-flavour QCD on lattices with temporal extent $N_\tau = 4$ and the chiral cumulant (b). Shown are results from calculations with different values of the quark mass on a $8^3 \times 4$ lattice.

Further scaling relations can be obtained by considering the quark mass dependence of the thermal susceptibility, $\chi_t \sim \partial\langle\bar{\psi}\psi\rangle/\partial t$, and the specific heat. A combined analysis of these allows a determination of the critical exponents $y_t$ and $y_h$. Results from the first analysis of this kind are summarized in Table 1 and compared with the known exponents for 3-$d$, $O(4)$ symmetric spin models [16] and mean field behaviour.

Table 1
Thermal ($y_t$) and magnetic ($y_h$) exponents from simulations of two-flavour QCD on an $8^3 \times 4$ lattice (LGT). Also given are mean field exponents (MF) and those for a 3-$d$, $O(4)$ symmetric spin model.

| | LGT | $O(4)$ | MF |
|---|---|---|---|
| $y_t$ | 0.69(7) | 0.446(5) | 0.5 |
| $y_h$ | 0.83(3) | 0.830(6) | 0.75 |

As can be seen from the Table the exponent $y_h$ extracted from lattice calculations is in very good agreement with the expected $O(4)$ model behaviour. It corresponds to a value of 0.205(45) for the exponent $1/\delta$, which is significantly below the value 1/3 expected for a mean field critical exponent. The exponent $y_t$, however, comes out too large in the present calculation. It may be expected that the determination of $y_t$ is more sensitive to finite volume effects than that of $y_h$. In the limit of vanishing quark masses it is $y_t$ that controls the finite volume scaling behaviour of thermodynamic quantities.

Clearly the analysis of critical exponents on a $8^3 \times 4$ lattice can only be considered as a first exploratory step towards a detailed investigation of critical exponents in two-flavour QCD. In the near future calculations on larger lattices with smaller quark masses will certainly lead to much better determinations of critical exponents. This should allow to distinguish between mean field and $O(4)$ critical behaviour.

## 4. CHIRAL BEHAVIOUR CLOSE TO $T_c$

Effective theories, deduced from QCD in order to describe low energy properties of hadrons, establish a close link between hadronic properties and the non-perturbative structure of the QCD vacuum, which is described by various non-vanishing condensates. This is, for instance, used in the operator product expansion (OPE) for correlation functions of hadronic currents [17]. It allows to relate hadron masses to condensates of the quark and gluon fields. The temperature and density dependence of the latter has been discussed within the context of chiral perturbation theory [7]. The applicability of the OPE as well as chiral perturbation theory is limited to low temperatures. In the vicinity of the QCD phase transition from hadronic matter to the quark gluon plasma phase, however, the hadronic medium becomes so dense that the influence of many particle states and resonances can no longer be neglected. In this regime an entirely non-perturbative approach becomes mandatory.

The numerical results for the energy density discussed in Section 2 suggest that up to temperatures of about $0.9T_c$ the hadronic medium is well described by a gas of hadrons, in which only the lightest states contribute. The slow variation of the energy density and pressure below $0.9T_c$ shows that also the gluon condensate will vary only little with temperature. This is also expected from chiral perturbation theory, which suggests that the temperature dependence of $\langle G^2 \rangle_T$ starts only at $O(T^8)$ [7]. For the chiral condensate $\langle \bar\psi\psi \rangle$ chiral perturbation theory gives, however, a stronger dependence on temperature,

$$\langle \bar\psi\psi \rangle(T) = \langle \bar\psi\psi \rangle(0) \Big[ 1 - \frac{n_f^2-1}{n_f} \Big( \frac{T^2}{12 f_\pi^2} \Big) - \frac{n_f^2-1}{2n_f^2} \Big( \frac{T^2}{12 f_\pi^2} \Big)^2 + O(T^6) \Big] \quad . \tag{11}$$

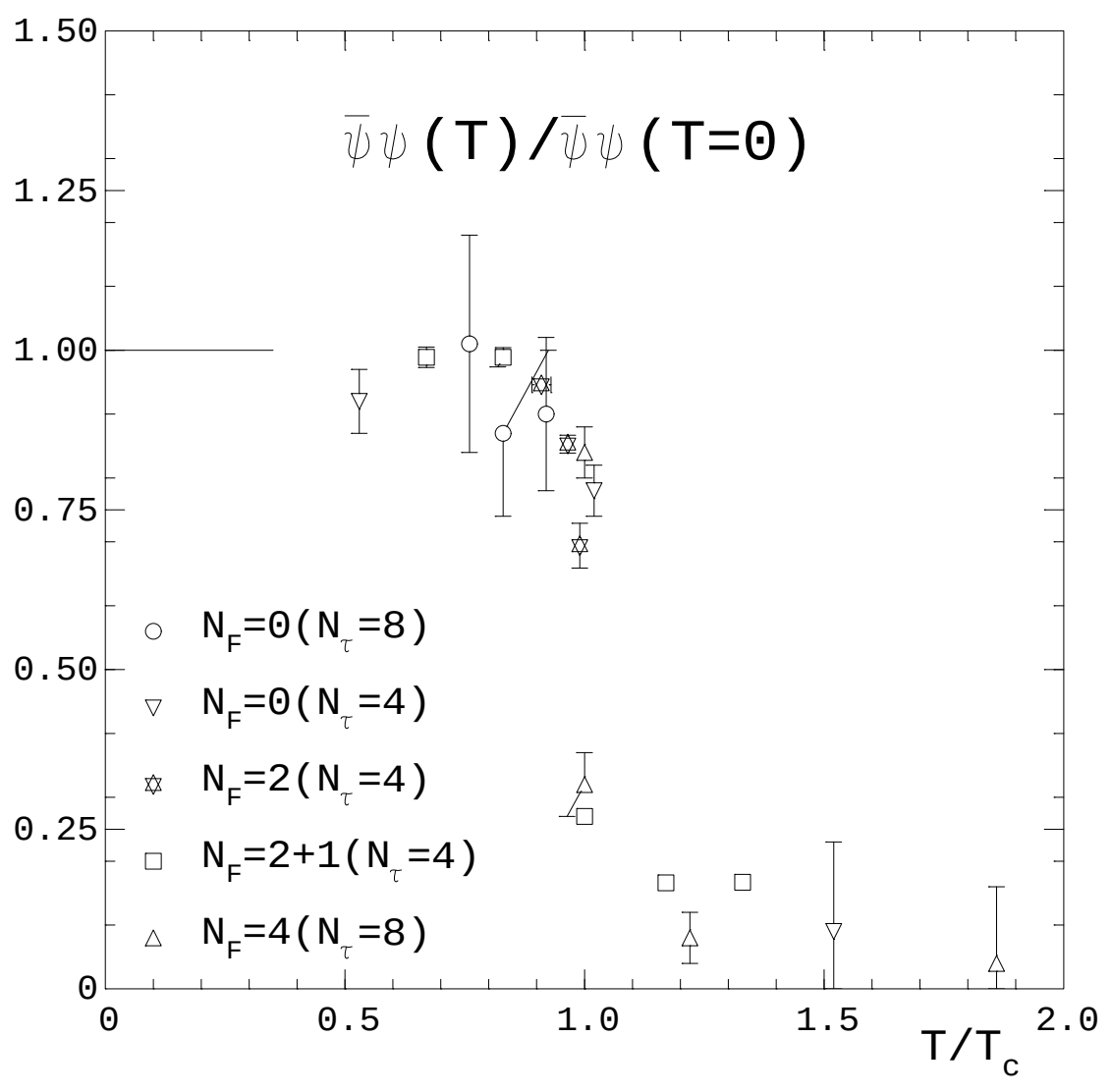

Figure 8. The normalized chiral condensate versus $T/T_c$. Shown are results from simulations with various numbers of flavours. Only in the pure gauge case ($n_f = 0$) have the results been extrapolated to vanishing quark mass. The zero quark mass extrapolation for QCD with dynamical quarks may still lead to a somewhat stronger temperature dependence.

In Figure 8 we show a collection of results for $\langle \bar\psi\psi \rangle$ obtained from lattice calculations for QCD with various number of flavours. They suggest at most a weak dependence of the chiral condensate on temperature, which may already indicate that we cannot expect to find drastic modifications of hadronic properties at least up to temperatures $T \sim 0.9 T_c$.

Information about hadron masses and decay constants can be extracted in lattice simulations from the long-distance behaviour of correlation functions of hadron operators

$$G_H(x) = \langle H(x) H^+(0) \rangle \rightarrow e^{-m_H |x|} \quad , \quad x \equiv (\tau, \vec{x}) \quad , \tag{12}$$

with $H(x)$ denoting an operator with the appropriate quantum numbers of the hadronic state under consideration, for instance $H(x) = \bar\psi_u(x)\gamma_5\psi_d(x)$ for the pion. At zero temperature one studies the behaviour of the correlation function at large Euclidean times $\tau$. From the exponential decay of $G_H$ at large values of $\tau$ one deduces the mass of the lightest state in this channel, whereas the amplitude is related to the decay constant of such a hadronic state. At finite temperature the Euclidean time extent is limited, $0 \le \tau \le 1/T$. It thus is preferable to study the behaviour of $G_H$ for large spatial separations, $|\vec{x}| \rightarrow \infty$.

This yields information about the finite temperature screening masses which are related to pole masses as long as there is a bound state in the quantum number channel under consideration.

Finite temperature screening masses have been studied in lattice simulations of QCD for quite some time [18]. A very drastic qualitative change in the screening masses is seen when one crosses the QCD transition temperature. Parity partners become degenerate above $T_c$, the pseudo-scalar mass becomes massive and approaches the free quark/anti-quark value, $m_{\rm meson} = 2\pi T$, at large temperatures. These features do not seem to depend much on the number of quark flavours. In particular, they also have been found in quenched QCD simulations. It thus seems to be meaningful to first study the behaviour of hadronic properties in the quenched approximation where results on large lattices with high accuracy can be obtained. In the following we will describe the results of such an investigation performed on a rather large lattice ($32^3 \times 8$) close to the phase transition temperature [19].

### 4.1. The GMOR Relation and the Pion Decay Constant

The GMOR relation relates the chiral condensate to the pion mass and pion decay constant,

$$f_\pi^2 m_\pi^2 = m_q \langle\bar{\psi}\psi\rangle_{m_q=0} \quad . \tag{13}$$

Below $T_c$ the pion is a Goldstone particle, its mass squared depends linearly on the quark mass,

$$m_\pi^2 = a_\pi m_q \quad . \tag{14}$$

A calculation of the chiral condensate and the pion mass at different values of the quark mass allows the determination of the pion slope, $a_\pi$, and the zero quark mass limit of the condensate, $\langle\bar{\psi}\psi\rangle_{m_q=0}$.

The pion decay constant $f_\pi$ can be determined directly from the relevant matrix element

$$\sqrt{2} f_\pi m_\pi^2 = \langle 0|\bar{\psi}_u \gamma_5 \psi_d|\pi^+\rangle \quad . \tag{15}$$

The square of the matrix element appearing on the right hand side of Eq. 15 is proportional to the amplitude of the pion correlation function and can be determined in a Monte Carlo calculation. A comparison of $f_\pi$ determined this way with the ratio $\langle\bar{\psi}\psi\rangle_{m_q=0}/a_\pi$ thus provides a direct test of the GMOR relation at finite temperature. This is shown in Figure 9a at $T \simeq 0.92 T_c$. Moreover, we can compare the value of $f_\pi$ calculated at finite temperature with corresponding zero temperature results. This is shown in Figure 9b. Clearly we neither have evidence for violations of the GMOR relation nor for a significant change of $f_\pi$ with temperature below $T_c$.

The sudden change of "$f_\pi$" above $T_c$ reflects the drastic change in the structure of the pseudo-scalar correlation function. The latter does no longer have a pole, which would correspond to a Goldstone-particle (pion). The existence of a pseudo-scalar bound state above $T_c$ thus is questionable. Certainly for large temperatures such a state does not exist, the correlation function is dominated by a quark/anti-quark cut. The notion of $f_\pi$ used for the square root of the amplitude of the correlation function should thus be used with caution above $T_c$.

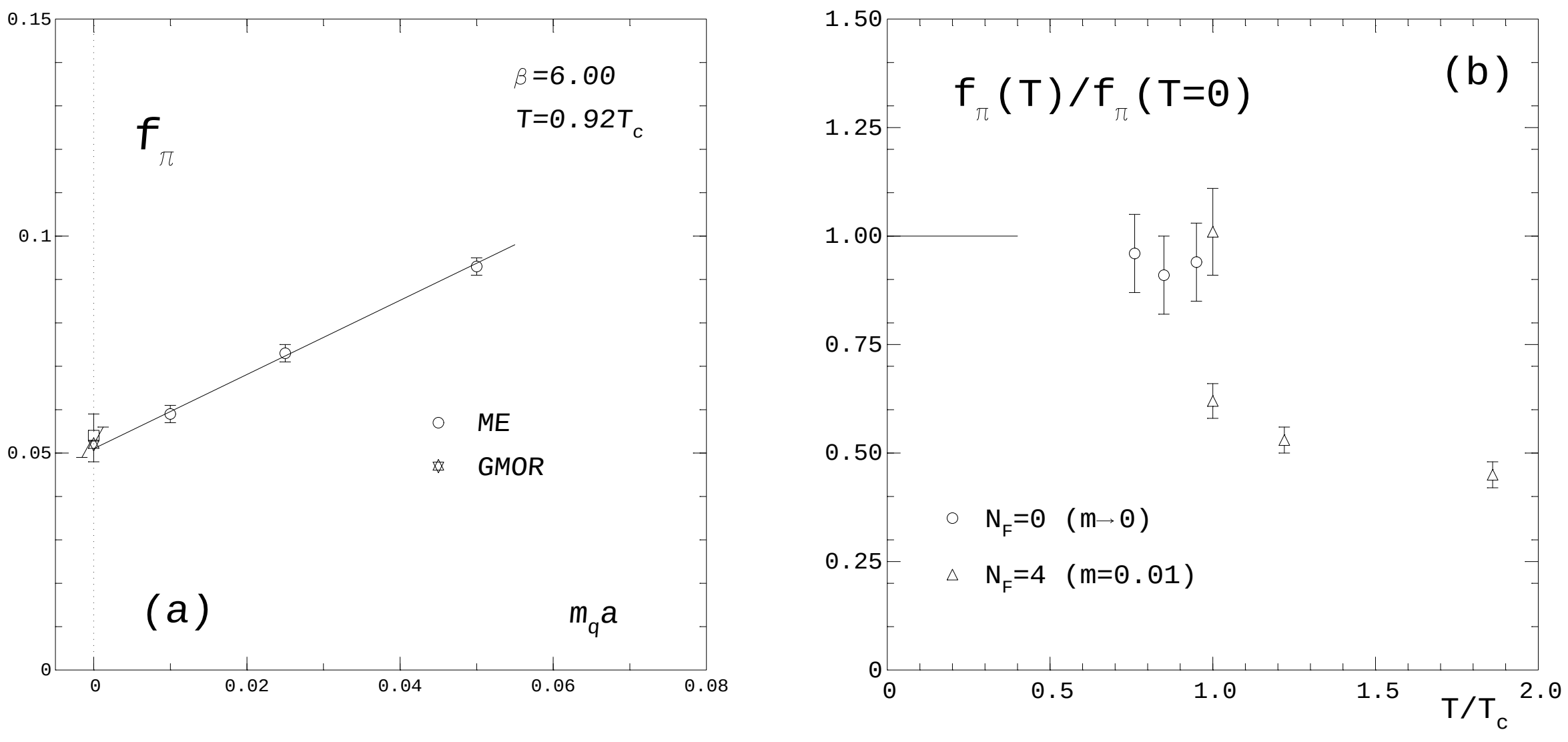

Figure 9. The pion decay constant at $T = 0.92T_c$ as a function of the quark mass (a) and its temperature dependence (b). In Figure (a) we show a comparison of results in quenched QCD ($n_f = 0$) obtained from the amplitude of the pion correlation function through Eq.(15) (circles). The zero quark mass extrapolation is compared with the result obtained from an application of the GMOR relation (star). Also shown is the zero temperature result at this value of the gauge coupling (square).

## 4.2. Vector Meson Mass and Nucleon Mass

The possibility of a variation of the $\rho$ meson mass with temperature is particularly interesting because of its possible experimental consequences. It might lead to modifications of dilepton spectra, which are experimentally detectable. The temperature dependence of the meson masses has been discussed within the framework of the OPE. Arguments have been given that the meson masses are temperature independent up to $O(T^2)$. The Monte Carlo calculations of the vector meson correlation function at finite temperature also show no significant temperature dependence of the mass even close to $T_c$. In Figure 10a we show the result of our calculation of the screening mass in the vector channel correlation function at $T \simeq 0.92T_c$ and compare this with zero temperature calculations at the same value of the gauge coupling. There is no evidence for any temperature dependence. The same holds true for the nucleon mass, although the details are more subtle in this case. As can be seen in Figure 10b there is a clear difference between the local masses, $m_N(z) \sim \ln G_N(z)/G_N(z+1)$, extracted on a large zero temperature lattice and those extracted at finite temperature on a lattice of size $32^3 \times 8$. However, as the nucleon is a fermion, there is a non-negligible contribution from the non-zero lowest Matsubara mode to the nucleon screening mass,

$$m_N = \sqrt{\tilde{m}_N^2 + (\pi T)^2} \qquad (16)$$

After removing the thermal contribution, $\pi T$, the nucleon mass, $\tilde{m}_N$, agrees with the zero temperature result within statistical errors.

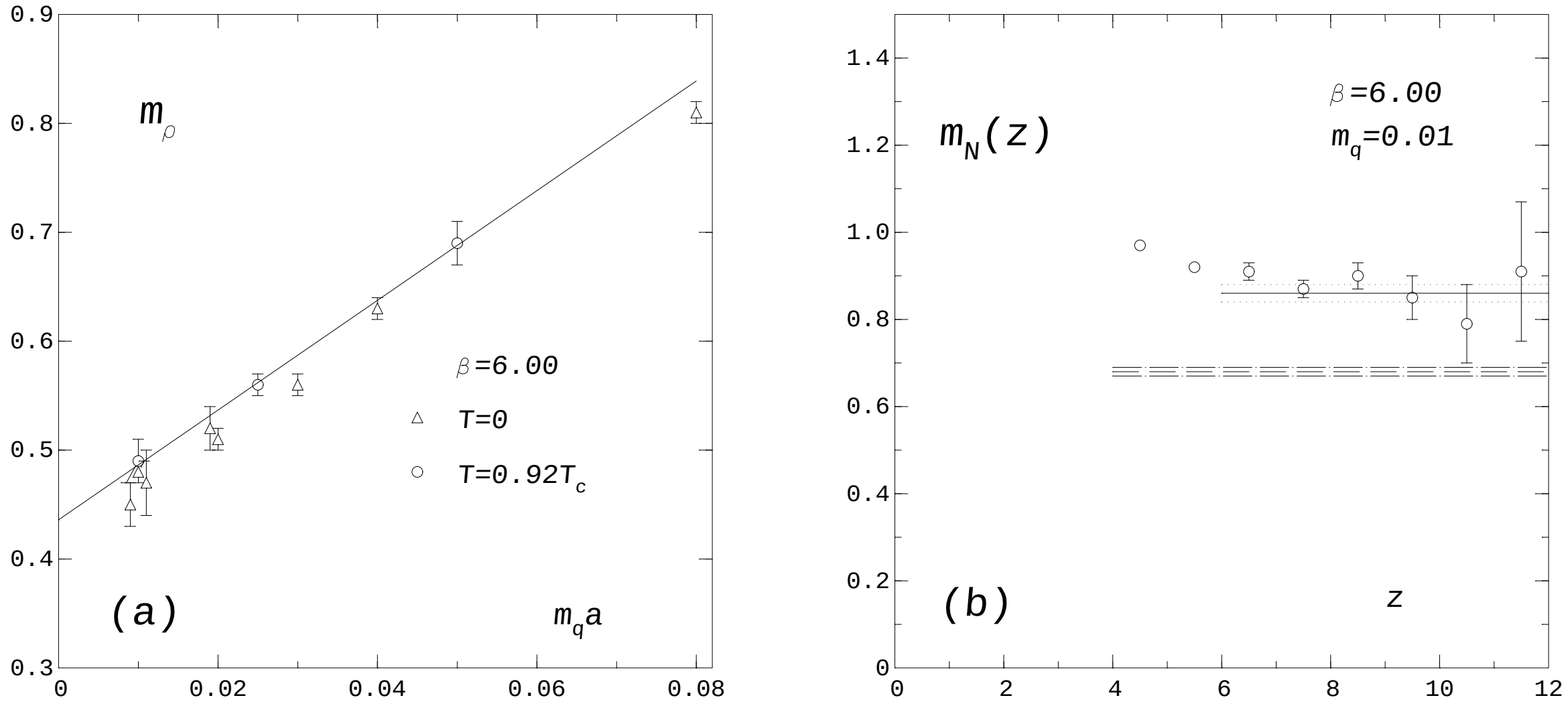

Figure 10. Temperature dependence of hadron masses: In Figure (a) we show a comparison of $\rho$-mass calculations at zero and non-zero temperature, $T = 0.92T_c$ for a fixed value of the gauge coupling, $6/g^2 = 6.0$ and for various values of the quark mass. In Figure (b) estimates for the nucleon mass obtained from the nucleon correlation function at distance $z$ are shown. These local masses converge to the nucleon mass in the limit $z \to \infty$. The horizontal band indicates the corresponding extrapolation at zero temperature. (See text for appropriate subtraction of the finite temperature Matsubara mode).

## 5. CONCLUSIONS

We have discussed three topics in finite temperature QCD: the structure of the equation of state, the order of the chiral phase transition and the dependence of hadron properties on temperature. A detailed quantitative understanding of these issues plays an important role in our attempts to clarify the structure of QCD and its non-perturbative infrared behaviour at high temperature.

Studies of the equation of state for pure gauge theories have now reached an accuracy, where systematic effects in the lattice approximations can be controlled and an extrapolations to the continuum limit can be performed with some confidence. We can expect that similarly accurate results will be available for QCD with dynamical fermions in the very near future. This then will provide important input for the description of the hydrodynamic expansion of hot hadronic matter created in heavy ion collisions.

The numerical analysis of the chiral transition will help to understand the relevant massless degrees of freedom in the vicinity of $T_c$, the role of composite scalar fields for the dynamics of the phase transition and the interplay between the deconfinement and chiral symmetry restoration mechanisms. First steps in this direction have now been undertaken through quantitative studies of critical exponents for two-flavour QCD.

A calculation of the temperature dependence of hadron properties is of immediate relevance for the discussion of possible experimental signals for the QCD phase transition in heavy ion experiments. Any change of hadron masses, their widths and decay constants should lead to experimentally observable consequences. Although drastic changes of the particle spectrum have been observed in lattice calculations, so far no sign of a temperature dependence has been found below $T_c$. This may be due to the current limitations for the study of hadron properties at finite temperature (small lattices, large quark masses, quenched approximation). This issue certainly will be studied in much more detail with more realistic parameters in the future.